\documentclass[iop]{emulateapj}

\usepackage{atbegshi}

\def\ltsima{$\; \buildrel < \over \sim \;$}
\def\simlt{\lower.5ex\hbox{\ltsima}}
\def\gtsima{$\; \buildrel > \over \sim \;$}
\def\simgt{\lower.5ex\hbox{\gtsima}}
\def\kpc{{\rm\,kpc}}

\def\lsun{{\rm\,L_\odot}}

\def\hst{{\it HST}}

\def\deg{^\circ}

\def\s{\ifmmode \widetilde \else \~\fi}
\def\={\overline}

\def\spose#1{\hbox to 0pt{#1\hss}}

\def\lta{\mathrel{\spose{\lower 3pt\hbox{$\mathchar"218$}}
     \raise 2.0pt\hbox{$\mathchar"13C$}}}
\def\gta{\mathrel{\spose{\lower 3pt\hbox{$\mathchar"218$}}
     \raise 2.0pt\hbox{$\mathchar"13E$}}}
\def\Dt{\spose{\raise 1.5ex\hbox{\hskip3pt$\mathchar"201$}}}    
\def\dt{\spose{\raise 1.0ex\hbox{\hskip2pt$\mathchar"201$}}}    

\def\dotsfill{\leaders\hbox to 1em{\hss.\hss}\hfill}

\def\FeH{{\rm[Fe/H]}}

\shorttitle{}
\shortauthors{N. F. Martin et al.}

\begin{document}


\title{A rogues gallery of Andromeda's dwarf galaxies I. A predominance of red horizontal branches}


\author{Nicolas F.\ Martin$^{1,2}$, Daniel R.\ Weisz$^3$, Saundra M.\ Albers$^3$, Edouard Bernard$^4$, Michelle L.\ M.\ Collins$^5$, Andrew E.\ Dolphin$^6$, Annette M.\ N.\ Ferguson$^7$, Rodrigo A.\ Ibata$^1$, Benjamin Laevens$^8$, Geraint F.\ Lewis$^9$, A.\ Dougal Mackey$^{10}$, Alan McConnachie$^{11}$, R.\ Michael Rich$^{12}$, Evan D.\ Skillman$^{13}$}

\email{nicolas.martin@astro.unistra.fr}

\altaffiltext{1}{Universit\'e de Strasbourg, CNRS, Observatoire astronomique de Strasbourg, UMR 7550, F-67000 Strasbourg, France}
\altaffiltext{2}{Max-Planck-Institut f\"ur Astronomie, K\"onigstuhl 17, D-69117 Heidelberg, Germany}
\altaffiltext{3}{Department of Astronomy, University of California, Berkeley, 94720, USA}
\altaffiltext{4}{Universit\'e C\^ote d'Azur, OCA, CNRS, Lagrange, France}
\altaffiltext{5}{Department of Physics, University of Surrey, Guildford, GU2 7XH, Surrey, UK}
\altaffiltext{6}{Raytheon, 1151 E. Hermans Road, Tucson, AZ 85756, USA}
\altaffiltext{7}{Institute for Astronomy, University of Edinburgh, Royal Observatory, Blackford Hill, Edinburgh EH9 3HJ, UK}
\altaffiltext{8}{Institute of Astrophysics, Pontificia Universidad Cat—lica de Chile, Av. Vicu–a Mackenna 4860, 7820436 Macul, Santiago, Chile}
\altaffiltext{9}{Sydney Institute for Astronomy, School of Physics, A28, The University of Sydney, NSW 2006, Australia}
\altaffiltext{10}{Research School of Astronomy and Astrophysics, Australian National University, Canberra, ACT 2611, Australia}
\altaffiltext{11}{National Research Council, Herzberg Institute of Astrophysics, 5071 West Saanich Road, Victoria, BC V9E 2E7, Canada}
\altaffiltext{12}{Department of Physics and Astronomy, University of California, Los Angeles, PAB, 430 Portola Plaza, Los Angeles, CA 90095-1547, USA}
\altaffiltext{13}{Minnesota Institute for Astrophysics, University of Minnesota, Minneapolis, MN 55441, USA}

\begin{abstract}
We present homogeneous, sub-horizontal branch photometry of twenty dwarf spheroidal satellite galaxies of M31 observed with the \emph{Hubble Space Telescope}. Combining our new data for sixteen systems with archival data in the same filters for another four, we show that Andromeda dwarf spheroidal galaxies favor strikingly red horizontal branches or red clumps down to $\sim10^{4.2}\lsun$ ($M_V\sim-5.8$). The  age-sensitivity of horizontal branch stars implies that a large fraction of the M31 dwarf galaxies have extended star formation histories (SFHs), and appear inconsistent with early star formation episodes that were rapidly shutdown. Systems fainter than $\sim10^{5.5}\lsun$ show the widest range in the ratios and morphologies of red and blue horizontal branches, indicative of both complex SFHs and a diversity in quenching timescales and/or mechanisms, which is qualitatively different from what is currently known for faint Milky Way (MW) satellites of comparable luminosities. Our findings bolster similar conclusions from recent deeper data for a handful of M31 dwarf galaxies. We discuss several sources for diversity of our data such as varying halo masses, patchy reionization, mergers/accretion, and the environmental influence of M31 and the Milky Way on the early evolution of their satellite populations. A detailed comparison between the histories of M31 and MW satellites would shed signifiant insight into the processes that drive the evolution of low-mass galaxies. Such a study will require imaging that reaches the oldest main sequence turnoffs for a significant number of M31 companions.

\end{abstract}


\section{Introduction}
Our knowledge of the lowest-mass galaxies primarily comes from observations of our immediate surroundings. Due to their intrinsic faintness, Local Group dwarf galaxies tend to be the only low-mass systems that can be studied in great detail via their resolved stellar populations \citep[e.g.,][and references therein]{mateo98,tolstoy09,mcconnachie12}. Their close proximity does not come without difficulties as, for instance, the properties of these low-mass stellar systems may be greatly influenced by their environment. The most direct example of this effect is the gas content of dwarf galaxies that depends on distance to the host or the Local Group in general \citep[e.g.,][]{grcevich09, spekkens14} and may be an indication of the transformative aspects of orbiting too close to one's host \citep[e.g.,][]{mayer01a}. 

Within the Local Group, the current observational situation for low-mass galaxies is far from ideal. Due to their close proximity, the bulk of observational efforts have been aimed at satellites of the Milky Way (MW). Thus, our broad understanding of low-mass galaxies, including star formation histories, chemical evolution, dark matter content, and their use as benchmarks for cosmological simulations, are based on a set of galaxies that reside in a common galactic ecosystem \citep[e.g.,][]{simon07,tolstoy09, kirby11, boylan-kolchin12, brooks13, brown14, weisz14, wetzel16}. It remains unknown how well the MW satellite population reflects the general properties of low-mass spheroidal satellites, let alone low-mass galaxies in general.

Considerable efforts have been made to remedy this shortcoming, primarily through observations of M31 satellites and isolated dwarf galaxies \citep[e.g.,][]{leaman13,cole14,gallart15,skillman17}. We now have broad knowledge of the dark matter contents of most M31 satellites \citep[e.g.,][]{tollerud13,collins14} and a steadily increasing set of systems with spectroscopic metallicity and heavier element abundance measurements \citep[e.g.,][]{ho12,collins13,vargas14,ho15}.

Critically lacking, however, is a detailed knowledge of the lifetime evolution of the M31 satellites. While dedicated imaging surveys such as the Pan-Andromeda Archaeological Survey (PAndAS; \citealt{mcconnachie09}) have proven invaluable for discovery and structural characterization of M31 satellites \citep[e.g.,][]{martin16d}, the resulting color-magnitude diagrams (CMDs) are generally limited to the luminous, evolved phases of evolution (e.g., upper red giant branch (RGB) stars), which, because of the age-metallicity degeneracy, provide only coarse information about star formation histories (SFHs) over cosmologically relevant timescales \citep[e.g.,][]{aparicio09,weisz11,dolphin12}.

In order to precisely reconstruct the full SFHs of these faint and crowded populations of ancient main sequence turnoff (MSTO) stars at the distance of M31, it is necessary to conduct their observations with the \emph{Hubble Space Telescope} (\hst). To date, such deep observations are limited to $\sim$20\% of the known M31 systems: M32 (compact elliptical; \citealt{monachesi12}), NGC~147 and NGC~185 \citep[dwarf ellipticals;][]{geha15}, and six dwarf spheroidal galaxies, observed as part of the Initial Star formation and Lifetimes of Andromeda Satellites (ISLAndS) project \citep[And I, And II, And III, And XV, And XVI, And XXVIII;][]{weisz14b,monelli16,skillman17}. Beyond lacking counterparts to the elliptical systems, MW satellites of similar luminosity appear to have notably different SFHs from the ISLAndS systems. The six ISLAndS systems were quenched more than 5 Gyr ago (i.e., they contain few or no stars younger than 5 Gyr), at odds with MW systems like Carina, Fornax, or Leo~I despite being of similar luminosities and host distances \citep{skillman17}. Moreover, whereas MW satellites below $\sim10^{5.5}\lsun$ ($M_V\sim-9.0$) tend to harbor only old stars (e.g., Draco, Hercules; \citealt{weisz14}), the faintest system in the ISLAndS program, And~XVI ($10^{4.8\pm0.1}\lsun$; \citealt{martin16d}) has a SFH that extends for $\sim$8~Gyr, which challenges the notion that low-mass galaxies are ubiquitously quenched by reionization.  Shallower CMDs of other faint M31 companions are also suggestive of similarly extended SFHs (e.g., \citealt{weisz14}), reinforcing the conclusions of first generation photometric studies of the M31 satellites that noted the presence of unusually red horizontal branch (HB) populations \citep{dacosta96,dacosta00,dacosta02}, indicative of extended episodes of star formation in the early ages of the galaxies' evolution.

While there is clear value in extending the ISLAndS program to the rest of the M31 satellites, the necessary \hst\ observations require relatively long integration times. As a compromise, we have pursued an exploratory program that acquired sub-HB HST/ACS imaging of all Andromeda dwarf spheroidal companions known as of late 2014 with no existing HST data, except for And~XIX and XXVII whose surface brightness is too low to yield useful, shallow CMDs. The observations of these 16 systems, combined with archival data for another 4, represent the most comprehensive study of the stellar contents of Andromeda satellite galaxies, with which we can broadly assess the diversity of their evolution relative to their MW counterparts.

In this first paper in the series, we present the data and focus on the HB morphology of all systems observed by our program. Future papers will quantify what constraints the current data impart on the recent SFH of these system and refine their distance estimates from the location of their HBs. Here, we report a qualitative predominance of red HBs (RHBs) or red clumps (RC) in the Andromeda companions, affirming and extending the pioneering work of Da Costa and collaborators on And~I, II, and III. The paper is structured as follows: \S \ref{sec:obs} presents the observations, data reduction, and resulting data set; \S \ref{sec:analysis} analyzes the HB content of the 16 galaxies in our sample, complemented by archival data for 4 systems whose observations were conducted with the same filters. Finally, \S \ref{sec:discuss} discusses our results and the systematically red HBs of Andromeda dwarf galaxies down to $\sim10^{5.5}\lsun$, and for a subset of the even fainter systems.

\section{Observations \& Data Reduction}
\label{sec:obs}

The details of the observations used in this paper are listed in Table~\ref{summary} and the properties of the dwarf galaxies are listed in Table~\ref{properties}. Observations of our sample of 16 galaxies were taken with the Advanced Camera for Surveys \citep[ACS;][]{ford98} aboard the \hst\ between October 1, 2014 and July 7, 2015 as part of HST-GO-13699 (PI: N.\ Martin). The target dwarf spheroidal galaxies are And~IX, X, XIV, XVII, XX, XXI, XXII, XXIII, XXIV, XXV, XXVI, XXIX, Cas~II and III, Lac~I, and Per~I. Systems that already have archival HST data were not targeted (And~I, II, III, V, VI, VII, XI, XII, XIII, XV, XVI, XVIII, XXVIII) as the two datasets will eventually be merged. In this paper, for a homogenous comparison, we restrict ourselves to observations performed with the set of F606W and F814W filters and to similar photometric depths. Archival observations of And~XI, XII, XIII, and XVIII meet this criteria and we include them in our analysis.

\begin{figure*}
\begin{center}
\includegraphics[width=\hsize]{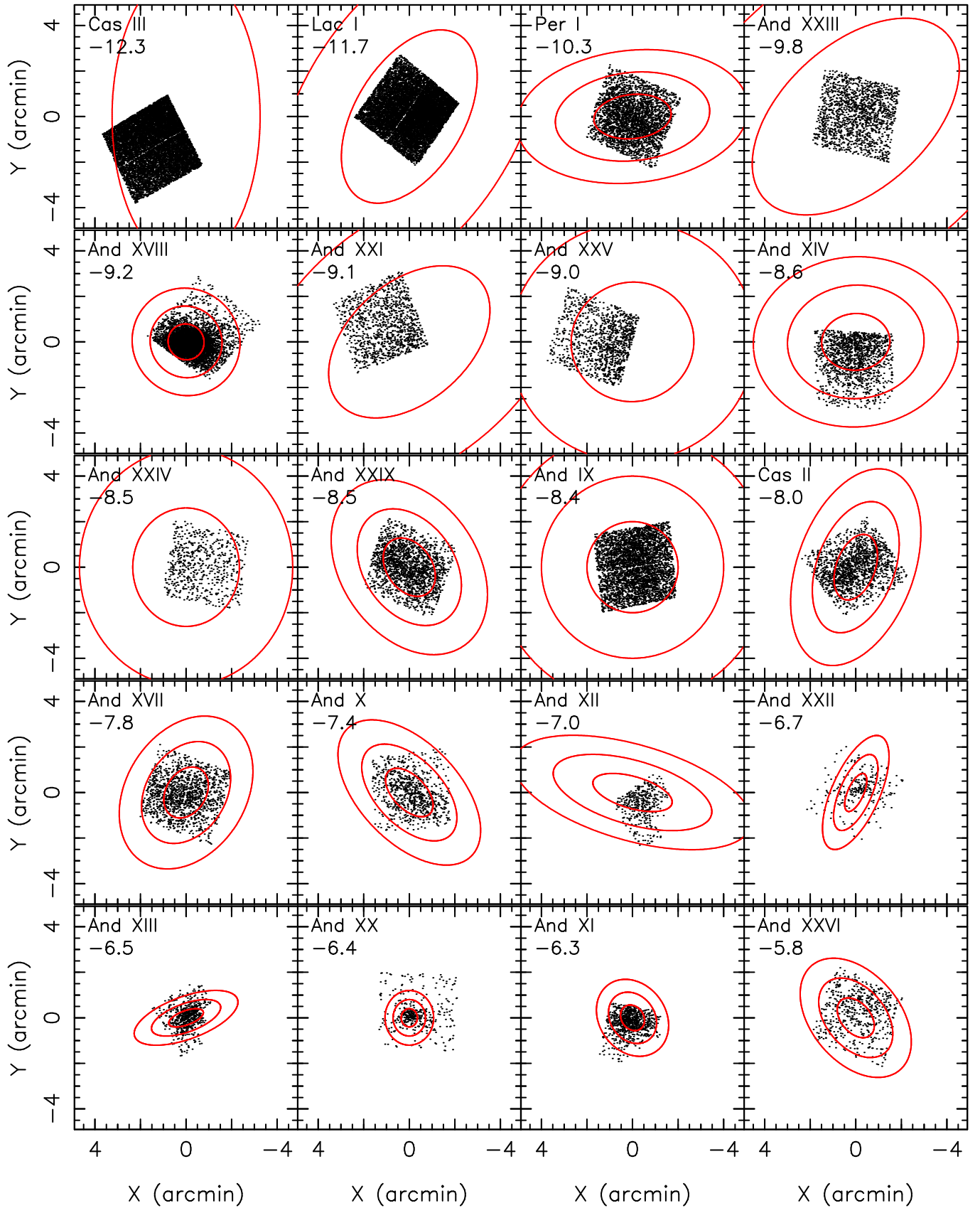}
\caption{\label{spatial}Spatial distribution of all star-like sources for the 20 dwarf galaxies in our sample. The concentric ellipses correspond to the $1r_h$, $2r_h$, and $3r_h$ regions as inferred by \citet{martin13a}, \citet{martin13c}, \citet{slater15}, \citet{martin16d}, and \citet{rhode17}. The systems are ordered by decreasing total luminosity and the total $V$-band magnitude, $M_V$, of each system is given in the top-left corner of each panel.}
\end{center}
\end{figure*}

A single field was observed for each galaxy of the HST-GO-13699 program. As much as possible, fields were chosen to target the center of a dwarf galaxy, but small offsets were sometimes necessary to avoid the presence of bright foreground stars. Figure~\ref{spatial} compares the distribution of detections consistent with being point-sources in the reduced data with ellipses delimiting the 1, 2, and $3 r_h$ regions of each galaxy. Obviously, the region sampled by ACS changes from system to system. It corresponds to smaller regions with at least 10--20 percent of a dwarf galaxy's content for the most extended systems (i.e., also the brightest; \citealt{brasseur11b}), but covers a significant fraction of the smallest dwarf galaxies, like And~XX and XXII (see the fourth column of Table~\ref{properties}).

For each field, a single orbit was split between two filters, F606W and F814W, with total integration times for each system listed in Table \ref{summary}. Observations in each filter were split into two sub-exposures taken at the same position to mitigate the effects of cosmic rays. To maximize time for science exposures, we opted not to dither the observations, and thus we did not fill the chip gap. In parallel, we observed a single pointing with the Wide Field Camera 3 (WFC3) in F606W and F814W. In half of the cases, the $\sim$6\arcmin\ angular separation between ACS and WFC3 meant the WFC3 field only sampled the very outer region of the target galaxy, where the expected stellar densities from ground-based imaging were known to be vanishingly low \citep{martin16d}.

We performed point spread function (PSF) photometry on each of the \texttt{flc} images using \texttt{DOLPHOT}, a version of \texttt{HSTPHOT} \citep{dolphin00} that has been specifically modified for ACS and WFC3 observations. We used \texttt{DOLPHOT} with Tiny Tim PSF models and photometric parameters recommended by \citet{williams14}.

The archival data of And~XI, XII, and XIII were observed with the Wide-Field Planetary Camera 2 (WFPC2) during program HST-GO-11084 (PI: D.\ Zucker). These data were processed by \citet{weisz14} and we refer the reader to this paper for more detail. Finally, And~XVIII was observed with ACS as part of program HST-SNAP-13442 (PI: Tully) and used a very similar set-up to our program's, with $2\times1100$\,s per filter. These data were processed like those of our program.

For all datasets, photometric catalogs of detected objects were culled to include only well-measured stars. Specifically, to identify stars, we required: SNR$_{\mathrm{F606W}} > 5$, SNR$_{\mathrm{F814W}} > 5$, (sharp$_{\mathrm{F606W}}$ $+$ sharp$_{\mathrm{F814W} }$)$^2 < 0.1$, and (crowd$_{\mathrm{F606W}}$ $+$ crowd$_{\mathrm{F814W} }$)$ < 1.0$. The definition of each parameter is given in \citet{dolphin00} and \citet{dalcanton09}. To gauge the completeness of our CMDs, we performed $\sim$100,000 artificial star tests on three galaxies (Cas~III, Cas~II, And~XX) that span the range of galaxy luminosities and distances of our sample. In each case we found 50\% completeness limits to be 27.1$\pm0.1$ in F606W and 26.2$\pm0.2$ in F814W. At the apparent magnitudes of their HBs, we found the data were 85-90\% complete.

The photometry of each field is de-reddened using the extinction values provided by NED\footnote{\url{http://ned.ipac.caltech.edu/forms/calculator.html}} from the \citet{schlegel98} map, re-calibrated by \citet{schlafly11}.

 \section{Analysis}
 \label{sec:analysis}
 
\begin{figure*}
\begin{center}
\includegraphics[width=0.95\hsize]{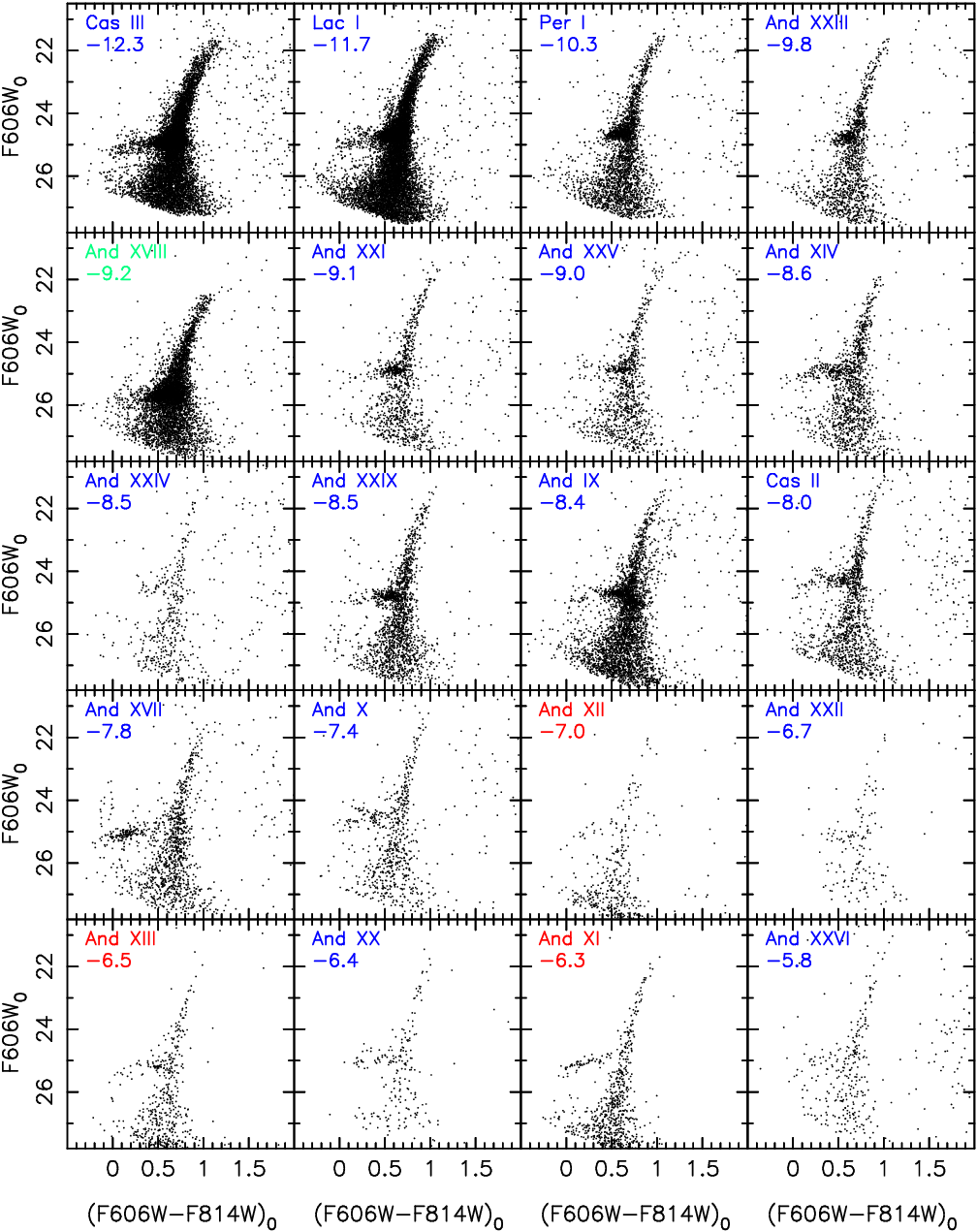}
\caption{\label{CMDs}CMDs of the 20 dwarf galaxies in our sample, ordered by decreasing luminosity. Each CMD includes stars within the full ACS field or the region with $2r_h$ of that dwarf galaxy, whichever is smallest. The numbers listed in the top-left corner of each panel correspond to the $M_V$ value of that system \citep{martin13a,martin13c,slater15,martin16d,rhode17}. Systems with blue labels were observed in our program (HST-GO-13699), whereas systems with red labels were observed with WFPC2 as part of program HST-GO-11084. And~XVIII, shown with a green label, was observed with ACS as part of the HST-SNAP-13442 program, which had similar observational properties. Note the predominance of red HBs overall.}
\end{center}
\end{figure*}

Figure~\ref{CMDs} presents the CMDs of the 20 dwarf galaxies in our sample. In order to minimize contamination from foreground stars, we restrict each CMD to the region within $2r_h$ of each dwarf galaxy, with $r_h$ values taken from \citet{martin13a}, \citet{martin13c}, \citet{slater15}, \citet{martin16d}, and \citet{rhode17}. The CMDs are ordered by decreasing galaxy luminosity. Variations in the density of the CMDs are a consequence of the galaxy's surface brightness and the fractional coverage of the ACS field. Contamination by M31 stellar halo stars is rarely an issue for these 20 dwarf galaxies that are relatively isolated in the M31 surroundings \citep[e.g.][]{martin13b}. A word of caution is needed for And~IX as it is projected on a substructure of the M31 stellar halo \citep{ibata14a}. Thus, its CMD is contaminated by a more metal-rich stellar population. This can be seen as stars redward of the RGB in Figure~\ref{CMDs}.

\begin{figure*}
\begin{center}
\includegraphics[width=\hsize]{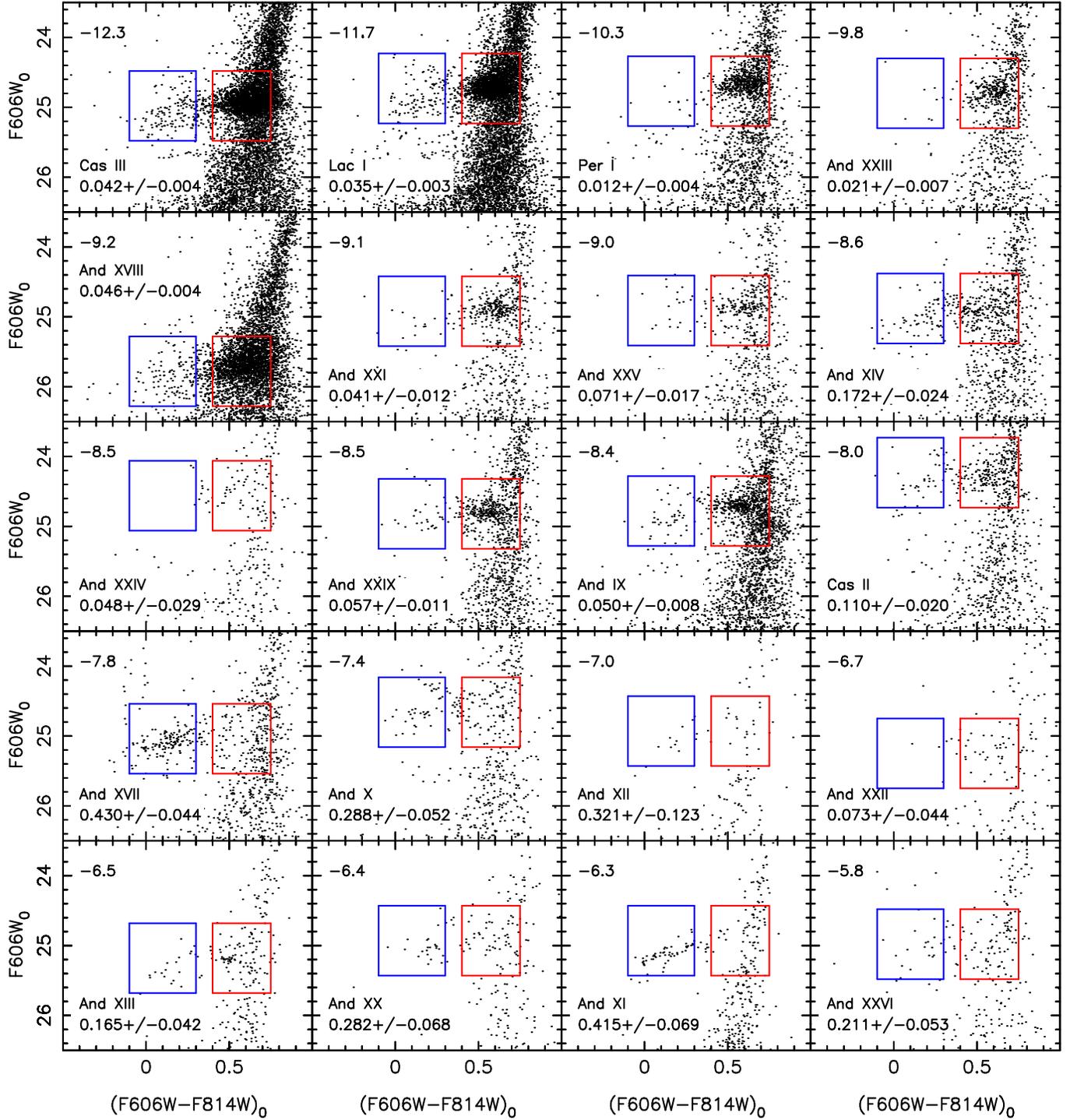}
\caption{\label{HB_CMDs}A zoom-in on the HB region of the CMDs presented in Figure~\ref{CMDs}. The red and blue boxes correspond to the selection boxes to count RHB/RC and BHB stars, respectively. Note that the red box purposefully extends red enough to include RC stars present in, e.g., Cas~III or Lac~I. By construction, this means that the red box also includes RGB stars. The boxes are fixed in color and allowed to move along the magnitude direction to track the change in distance from galaxy to galaxy. The number in the top-left corner of each panel is the $M_V$ value of the corresponding dwarf galaxy and the number listed in the bottom-left corner of each panel is the blue-to-total ratio, $\eta$. In the case of the more distant And~XVIII, this number is listed in the top-left corner of the panel.}
\end{center}
\end{figure*}

The most striking feature of the CMDs presented in Figure~\ref{CMDs} is the presence of a clear \emph{red} HBs or even RCs in most galaxies. This is even more evident in the CMDs of Figure~\ref{HB_CMDs} that are zoomed-in on the HB region. Even a very faint system like And~XIII ($M_V=-6.5^{+0.7}_{-0.5}$; \citealt{martin16d}) has a clear population of RHB stars, which is unlike MW satellites of the same luminosity (e.g., Hercules, M$_V \sim -6.6$) that have exclusively blue HBs (BHBs; \citealt{brown14}). Only And~XVII and XI show predominantly blue HB and And~XIV hosts a more balanced HB. The situation is more ambiguous for the faint systems And~X, XII, XIII, XX, XXII, XX, and XXVI that do not contain many stars.

Following \citet{dacosta96,dacosta00,dacosta02}, we quantify the dominance of RHBs by simply counting stars in two selection boxes focusing on the RHB and BHB regions, as shown by the red and blue boxes in Figure~\ref{HB_CMDs}. The colors of these boxes remain the same for every system but we allow for shifts in magnitude to account for changes to the distances of the dwarf galaxies\footnote{Tying these magnitude shifts to differences in the distance moduli of the dwarf galaxies \citep[e.g.,][]{aconn12} proved unsatisfactory due to the large uncertainties of some of the distance measurements. It emphasizes the need to more accurately derive these distances, which is one of our main goals with these data.}. The boxes are purposefully kept wide in the magnitude direction so they generously include the HB stars and our counts are not biased by changes in the HB morphology or stellar variability\footnote{Such variable stars can be see in the brighter half of the blue HB box for Cas~III and Lac~I.}. Although we would have likely to determine the color limits of the boxes based on literature values, we could not find studies that relied on the filter set we use here. We therefore chose to place the boxes so the blue one encompasses the blue part of the HB in for systems with the bluest HB (And XVII or XI) and similarly for the red HB (e.g., Cas~III, Lac~I, or And~XVIII). 

The resulting ratios, $\eta = n_\mathrm{BHB}/(n_\mathrm{BHB}+n_\mathrm{RHB}$), of numbers of stars in the BHB ($n_\mathrm{BHB}$) and BHB$+$RHB ($n_\mathrm{RHB}+n_\mathrm{RHB}$) selection boxes\footnote{It should be noted that the red box is tailored to include red clump stars (e.g., those of Cas~III or Lac~I), which means that it also includes RGB stars. As such, the ratios of counts in the BHB-to-RHB boxes cannot become very large, but they nevertheless carry significant changes from galaxy to galaxy.} are listed in the panels of Figure~\ref{HB_CMDs} and confirm our initial impression: all the galaxies in the brighter half of the sample have $\eta<0.1$, except for And~XIV ($\eta=0.189\pm0.024$). For systems with $M_V\gta-8.0$ (i.e., $L_V\lta10^{5.1}\lsun$), the ratios are noisier but are also spread over a wider range, up to $0.412\pm0.042$ in the case of the system with the bluest HB, And~XVII. Nevertheless, even at these luminosities, one can still find systems with fairly red HBs (And~XXII with $\eta=0.141\pm0.050$ or And~XIII with $\eta=0.165\pm0.042$).

\begin{figure*}
\begin{center}
\includegraphics[width=0.5\hsize,angle=270]{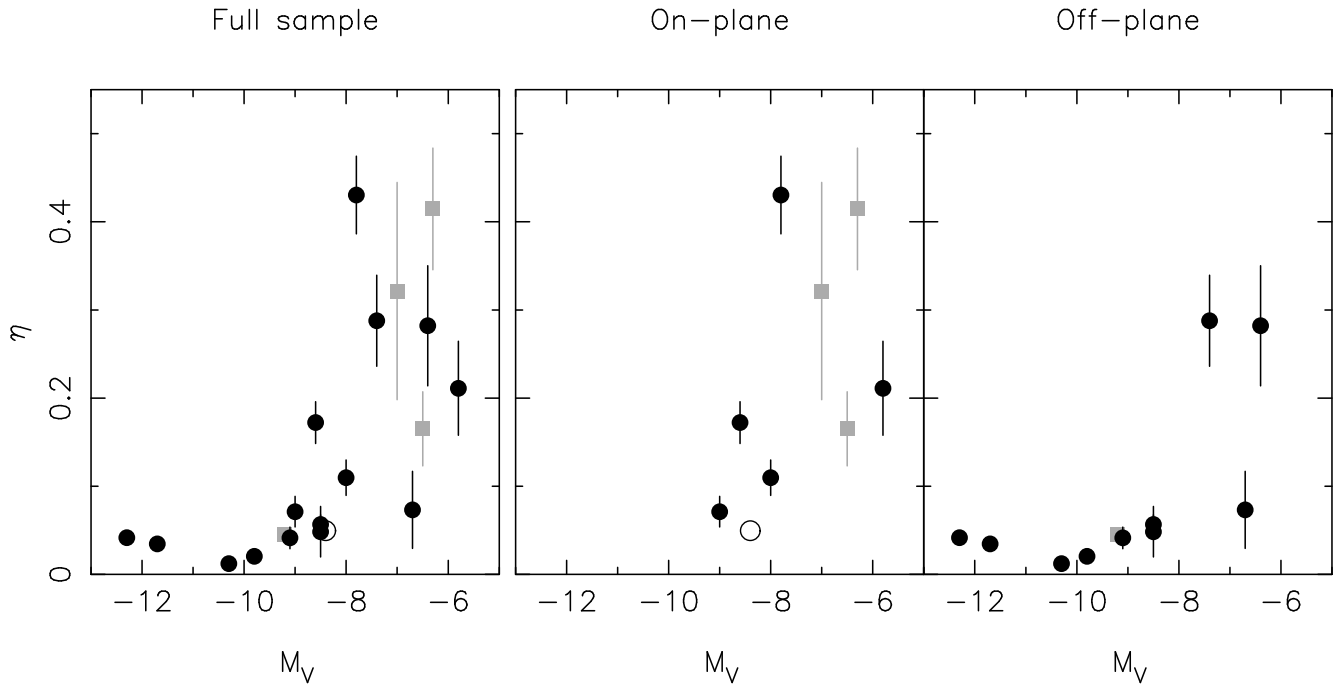}
\caption{\label{ratios}Ratio of numbers of blue-to-total HB stars, $\eta$, as a function of a dwarf galaxy's total luminosity. Gray symbols correspond to the 4 dwarf galaxies that were not part of our program and were taken from the archive (And~XI, XII, XIII, XVIII). The hollow dot represents And~IX, which is contaminated by a (metal-rich) M31 halo structure and whose $\eta$ is therefore biased low. $\eta$ is systematically low, indicative of red HBs, for systems brighter than $M_V\simeq-8.5$ but a transition happens at this luminosity. Dwarf galaxies with predominantly blue HBs (high $\eta$ values) exist below this limit but others are still dominated by red HBs.}
\end{center}
\end{figure*}

For a more comprehensive view of the variations in $\eta$, Figure~\ref{ratios} (left panel) presents its changes as a function of the magnitude of the galaxy. Broadly, two main effects are visible in this panel:

\begin{itemize}
\item There is a significant change in the color of HBs in the range $-9.0<M_V<-8.0$ or $L_V\sim10^{5.5}\lsun$. For brighter magnitudes, HBs are systematically measured to be red.
\item For fainter systems, the values of $\eta$ cover a much larger range, allowing for very blue HBs, but it is not a systematic trend and red HBs are still present.
\end{itemize}

For faint systems, shot noise becomes important and is responsible for the larger uncertainties on $\eta$. These are however not large enough to explain the changes in the ratio of blue-to-total HB stars. Contamination from field objects is also a concern but the WFC3 observations for the small dwarf galaxies And~XX, XXII, and XXVI can help us assess expected values of $\eta$ in the absence of member stars. These three dwarf galaxies are all small enough that their WFC3 fields should correspond to the typical field contamination. Furthermore, they cover a large range of spatial locations around M31 as they are spread over almost $20\deg$ in Galactic latitude between And~XXVI and And~XXII, located at $b=-14.7\deg$ and $-34.1\deg$, respectively. For the three systems, we measure $\eta=0.333\pm0.136$, $0.167\pm0.068$, and 0, respectively. More importantly, the $n_\mathrm{BHB}$ and $n_\mathrm{RHB}$ counts are extremely low in all cases ($<5$). We can therefore expect that the values calculated for the dwarf galaxies are not significantly contaminated as the counts are always larger than 19 in at least one of the boxes.

\begin{figure}
\begin{center}
\includegraphics[width=0.7\hsize,angle=270]{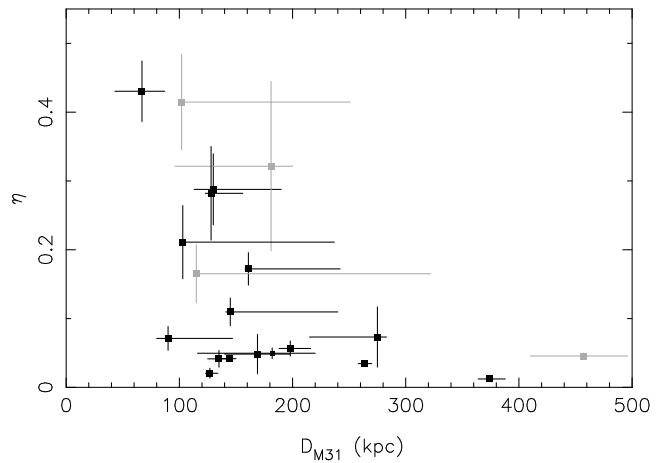}
\caption{\label{ratio_DM31}Changes in the ratio of the numbers of blue-to-total HB stars with the M31-centric distance of a dwarf galaxy. More distant systems disfavor blue HBs but surveys tend to be limited to bright dwarf galaxies in these regions. There is no obvious correlation between $\eta$ and $D_\mathrm{M31}$ in the more complete region within $\sim200\kpc$ of M31. M31-centric distances are taken from \citet{conn12}, \citet{martin13a}, \citet{martin13c}, \citet{slater15}, and \citet{rhode17}.}
\end{center}
\end{figure}

At this point, it is also important to investigate the impact of the M31-centric distance of a satellite on $\eta$. The gaseous content of dwarf galaxies is known to correlate with their distance from the host \citep[e.g.,][]{grcevich09} and distant dwarf galaxies, which have spent most of their orbital time far from their host, are also more prone to exhibit extended star formation histories \citep[e.g.,][]{weisz14}. Since the orbits of the studied dwarf galaxies are not known, the only proxy we can use is their current M31-centric distance. Figure~\ref{ratio_DM31} shows the changes of $\eta$ as a function of the M31-centric distance of the satellites. Although the most distant satellites disfavor blue HBs, one must keep in mind that the sensitivity of dwarf galaxy searches outside of the PAndAS footprint ($\sim150\kpc$ in projection) is such that discovered distant dwarf galaxies are more likely to be brighter and above the $\sim10^{5.5}\lsun$ transition visible in Figure~\ref{ratios} \citep[e.g.,][]{martin13a,slater15}. In the more complete $D_\mathrm{M31}\lta200\kpc$ region, we find no obvious impact of the M31-centric distance of the dwarf galaxy on $\eta$.

Differences in the aerial coverage of the various dwarf galaxies by the field of view is also worth discussing. It is well known that bright dwarf galaxies, around both the Milky Way \citep[e.g.,][]{tolstoy04} and Andromeda \citep[e.g.,][]{mcconnachie07a}, can host multiple stellar populations and that, when this is the case, redder HB populations are more centrally concentrated than their bluer counterpart \citep[e.g.,][]{tolstoy09}. Since the brighter dwarf galaxies of Andromeda are also the larger ones \citep{brasseur11b}, our observations probe these systems more centrally are, therefore, could be biased towards enhancing their red HB fraction. While this is certainly a concern, this does not appear to drive the dominance of red HBs. Indeed, if we are to assume very conservatively that all RHBs are restricted to the region within the half-light radius of the three systems with the smallest coverage (Cas~III/And~XXIII/And~XXI with fractional coverages of 0.12/0.19/0.20), $\eta$ would only change from 0.044/0.022/0.054 to 0.154/0.052/0.097 and remain in the lower part of Figure~\ref{ratios}. Since this is a very strong (and unrealistic) assumption, we conclude that the predominance of RBHs in Andromeda satellite dwarf galaxies is unlikely to be mainly driven by differences in the areal coverage of the data.

Beyond the study of the changes in the number of blue-to-total HB stars for the 20 dwarf galaxies in the sample, we also investigate potential differences in the dwarf galaxies that are aligned in the recently discovered thin disk of Andromeda satellites \citep{ibata13a}. The selection function of the sample, built from what was or not in the HST archive, explains the different luminosity range of the on- and off-plane samples (middle and right-hand panels of Figure~\ref{ratios}). Beyond this effect, there is no significant difference between the $\eta$ behavior of the two samples. This conclusion is in agreement with its ISLAndS counterpart \citep{skillman17} and in line with similar comparisons of the various properties of these systems by \citet{collins15}, who concluded against a significantly different formation and evolution route for the two sets of Andromeda satellites because of the absence of a significant difference between on- and off-plane dwarf-galaxy properties.

\section{Discussion}
\label{sec:discuss}
We have presented deep, sub-HB photometry of 20 dwarf galaxies satellite of M31 from new or archival HST data. These data sample two thirds of all known Andromeda satellite dwarf spheroidal galaxies and provide a gallery of homogeneous CMDs that can easily be compared over a wide luminosity range, from Cas~III ($10^{6.8}\lsun$) to And~XXVI ($10^{4.2}\lsun$). With but a few exceptions, these M31 dwarf galaxies display predominantly red HBs/RCs, even at the faint end of the sample. Only And~XVII ($10^{5.0}\lsun$) and And~XI ($10^{4.4}\lsun$) host distinctly blue HBs and a handful of galaxies show balanced HBs. In addition, there appears to be a transition in the HB content of the galaxies at $\sim10^{5.5}\lsun$ ($M_V\simeq-9.0$): only fainter than this limit do Andromeda's spheroidal satellites present a range of blue-to-total HB star ratios. However, even then, there remains faint systems whose HB is dominated by red stars (e.g., And~XIII, $10^{4.5}\lsun$).

Our data are not deep enough to enable a decomposition of the stellar populations at the oldest MSTO, as is the case in \citet{skillman17}. Thus, in this paper we only provide a qualitative assessment of the evolution of these systems, and we plan a quantitative SFH determination \citep[e.g.,][]{weisz14a} in a future paper in this series. Using our broad knowledge of stellar and dwarf galaxy evolution \citep[e.g.,][]{gallart05,tolstoy09} and \citet{skillman17} as a guide, we surmise that the presence of red HB and RC populations are due to a combined effect of age and metallicity evolution, which we refer to as extended SFHs, i.e., changes in both age and metallicity. The few reliable spectroscopic measurements of metallicities for the dwarf galaxies in our sample show that they likely are all, broadly speaking, metal-poor ($-2.3\lta\FeH\lta-1.5$), with maybe a hint of a luminosity--metallicity relation \citep{collins15} within the luminosity range of the sample. This could further compound the difficulty to derive a quantitative star-formation history from the morphology of the HB and explains why we focus here on a qualitative analysis. Beyond this, HB morphology is also known to be influenced by second parameter effects, e.g., helium abundance. For simplicity, we will restrict out discussion to age and metallicity effects, but recognize that a detailed interpretation of the HB morphology likely requires more than these two parameters \citep[e.g.,][]{gratton10}.

The galleries of Figures~\ref{CMDs} and \ref{HB_CMDs} suggest that most M31 dwarf galaxies have had a prolonged period of star formation, even down to $\sim10^{4.2}\lsun$.  Many formed a significant fraction their stars later, and likely at multiple metallicities, when compared to the oldest metal-poor globular clusters or dwarf galaxies that host purely blue HBs. It is clear that the prolonged period of star formation cannot by younger than $\sim$ 3--5 Gyr, otherwise our data would show the presence of young MS stars, which are not seen in any of the systems.

Our qualitative interpretation is bolstered by the more detailed and complementary study of a small sample of M31 dwarf galaxies by \citet{skillman17}, who reached similar conclusions, i.e., extended age and metallicity evolution, from the sub-MSTO observations of 6 dwarf galaxies. In particular, this ISLAndS sample shows that \emph{all} of the systems studied in detail, spanning $10^{4.8}<L<10^{6.5}\lsun$, exhibit an extended period of star formation and quenching at intermediate times, only 5--9 Gyr ago ($z\sim$1--1.5). Of particular interest, the faintest system in the ISLAndS sample, And~XVI, is the system that was quenched latest, which mirrors our discovery that even faint M31 dwarf galaxies can host predominantly red HBs (e.g., And~XIII and And~XXII). Analysis of the HB and variable star populations in the ISLAndS sample (Mart\'inez-V\'azquez et al. in preparation) shows that their 6 dwarf galaxies are systematically redder than that of their MW counterparts. They interpret this as an indication of more extended SFHs for the M31 systems. The shallower data we present here broaden this conclusion to include most of the M31 dwarf galaxies. Despite the difficulties of reliably modeling the HB region, \citet{makarova17} inferred from the same HST data we present here that And~XVIII hosts intermediate (2--8 Gyr) and old (12--14 Gyr) stars, further bolstering our qualitative assessment.

Evidence for systematically red HBs becomes less clear for galaxies fainter than $M_V$ $\sim-8$. Figure~\ref{ratios} shows $\eta$ values that range from $\sim0.05$ to $0.45$ for these fainter systems. This trend persists in excess of shot noise, suggesting that it is a physical effect. Taking $\eta$ as a coarse proxy for age, one interpretation of Figure \ref{ratios} is that the faint M31 satellites exhibit a wide range of SFHs. Those with larger values of $\eta$ have predominantly ancient SFHs, while those with lower values have more extended SFHs and intermediate-age stellar populations older than 3--5 Gyr. 

This is in contrast to MW satellites at similar luminosities (e.g., Hercules, Bo\"otes~I, Ursa~Major~I), all of which have purely ancient, metal-poor populations as indicated both by wide-field Subaru photometry \citep{okamoto12} and by SFHs measured from analysis of \hst-based CMDs that extend below the oldest main sequence turnoff \citep[e.g.,][]{brown14}. Affirmed by these MW satellites, the current picture of quenching in the lowest-mass galaxies is that the ultra-violet background associated with reionization truncated their star formation following the end of the reionization era \citep[e.g.,][]{bullock00, ricotti05, bovill09, tumlinson10}.

However, our suggestion of extended star formation and chemical evolution in some M31 satellites that should have quenched by $z\sim6$, via the scenario above, suggests a more complex picture. For example, the effects of reionization on low-mass galaxies may be more subtle than the abrupt quenching of star formation. A number of recent simulations suggest that reionization may more effectively prevent the accretion of fresh gas onto a low-mass halo, rather than photo-evaporate its existing gas supply, thus allowing low-level star formation to continue beyond the end of reionization \citep[e.g.,][]{onorbe15, fitts17}. Another possibility is that reionization may be non-uniform either owing to the statistical variations in some global background (e.g., patchy reionization) or due to the fact that different local sources (e.g., M31 vs. the MW vs. Virgo) were responsible for reionization \citep[e.g.,][]{busha10, lunnan12, ocvirk13, ocvirk16}. Such differences could be imprinted on the stellar populations of the faintest satellites. Finally, it could be that many of the M31 satellites with extended SFHs were hosted by more massive dark matter halos in the early Universe than MW systems of the same luminosity. A larger halo mass in the early Universe, would reduce the quenching effects of reionization on a given low-mass galaxy.

It is also possible that reionization is not the only mechanism that shapes the early evolution of very faint galaxies. For example, galaxy mergers or the re-accretion of gas at later times could lead to more complex stellar populations in faint systems \citep[e.g.,][]{deason14b}. In addition, the environmental influence of the MW and M31 on low-mass galaxies cannot be discounted. At present, there is little observational handle on the locations of MW or M31 satellites in the early Universe relative to their present-day hosts \citep[e.g.,][]{watkins13}. Thus, although the Local Group was quite large at early times \citep[$\sim350$ Mpc$^3$ at $z\sim7$ vs. $\sim 7$ Mpc$^3$ at $z=0$;][]{boylan-kolchin16}{}, it is possible that present-day satellites were close enough to the proto-MW or proto-M31 to be affected by their tides or ram pressure from their hot circumgalactic media. The influence of these processes could  have altered the evolutionary trajectory of any low-mass galaxy. 

While these conclusions are intriguing, they remain qualitative and speculative, and it would be very beneficial to solidify them with detailed and accurate SFHs of the 20 M31 dwarf galaxies presented here. This is one of our goals with this data set but, unfortunately, the limited depth of the data, which was driven by the necessity to keep this program competitive in terms of exposure time, will restrict the derived SFHs to only the last 5--8\,Gyr and will not reach the ancient epochs that can only be probed by the oldest MSTO. The mounting evidence that the MW and the M31 dwarf galaxies differ significantly should however encourage the community to enable an ISLAndS-like treatment of a large fraction (if not all) of the M31 dwarf galaxies. Variations in the evolutionary histories of M31 and MW satellites extend far beyond the history of two Local Group sub-systems. Indeed, these faint galaxies play an important roles from calibrating physics in detailed simulations to testing $\Lambda$CDM models of galaxy formation \citep[e.g.,][]{wetzel16} to constraining the shape of the faint-end of the high-redshift galaxy luminosity function \citep[e.g.,][]{boylan-kolchin15, weisz17}. However, most of our detailed understanding of dwarf galaxies currently stems from a small number of systems around the MW and one should be wary of mistaking the details of a specific system, on a specific orbit, around a specific host, for the global properties of dwarf galaxies in a group environment.

\begin{table*}
\caption{\label{summary}Listing of observations}
\begin{tabular}{l|lllll}
Galaxy Name & Program ID & Instrument & Exposure Time (s) & PI & Observing Dates\\
 & & & (F606W,F814W) &  &\\

\hline
And~IX & HST-GO-13699 & ACS & 1146,1146 & Martin & 2014-10-01, 2014-10-02\\
And~X & HST-GO-13699 & ACS & 1146,1146 & Martin & 2014-10-04\\
And~XI & HST-GO-11084 & WFPC2 & 19200,26400 & Zucker & 2007-09-06, 2007-09-09, 2007-09-17, 2007-09-20\\
And~XII & HST-GO-11084 & WFPC2 & 19200,26400 & Zucker & 2007-06-14, 2007-06-15, 2007-08-01, 2007-08-02\\
And~XIII & HST-GO-11084 & WFPC2 & 19200,26400 & Zucker & 2007-07-22, 2007-07-23, 2007-07-31, 2007-08-01, \\
 &  &  &  &  & 2007-08-02\\
And~XIV & HST-GO-13699 & ACS & 1110,1109 & Martin & 2014-11-28\\
And~XVII & HST-GO-13699 & ACS & 1146,1146 & Martin & 2015-07-02, 2015-07-03\\
And~XVIII & HST-SNAP-13442 & ACS & 1100,1100 & Tully & 2013-10-20\\
And~XX & HST-GO-13699 & ACS & 1128,1128 & Martin & 2014-12-07\\
And~XXI & HST-GO-13699 & ACS & 1148,1146 & Martin & 2015-03-02\\
And~XXII & HST-GO-13699 & ACS & 1110,1109 &Martin & 2015-02-03\\
And~XXIII & HST-GO-13699 & ACS & 1128,1128 & Martin & 2015-07-01\\
And~XXIV & HST-GO-13699 & ACS & 1168,1168 & Martin & 2015-07-01\\
And~XXV & HST-GO-13699 & ACS & 1168,1168 & Martin & 2015-07-03\\
And~XXVI & HST-GO-13699 & ACS & 1168,1168 &Martin & 2015-07-03\\
And~XXIX & HST-GO-13699 & ACS & 1118,1117 &Martin & 2014-12-22\\
Cas~II & HST-GO-13699 & ACS & 1168,1168 & Martin & 2014-11-12\\
Cas~III & HST-GO-13699 & ACS & 1194,1193 & Martin & 2015-03-03\\
Lac~I & HST-GO-13699 & ACS & 1148,1146 & Martin & 2015-07-06\\
Per~I & HST-GO-13699 & ACS & 1148,1146 & Martin & 2015-02-25\\
\end{tabular}
\end{table*}

\begin{table*}
\caption{\label{properties}Properties of the dwarf spheroidal galaxies}
\begin{tabular}{l|llllll}
Dwarf galaxy & Magnitude\footnote{Magnitude and distance measurements are taken from \citet{aconn12}, \citet{martin13a}, \citet{martin13c}, \citet{slater15}, \citet{martin16d}, and \citet{rhode17}.} & $D_\mathrm{M31}^\mathrm{a}$ (kpc) & Fractional coverage\footnote{This coverage corresponds to the integral of a dwarf galaxy's density profile over the HST fields.} & $n_\mathrm{BHB}$ & $n_\mathrm{RHB}$ & $\eta$ \\
\hline
And~IX & $-8.5\pm0.3$ & $182^{+38}_{-66}$ & 0.50 & 42 & 805 & $0.050\pm0.008$\\
And~X & $-7.4\pm0.3$ & $130^{+60}_{-17}$ & 0.82 & 40 & 99 & $0.288\pm0.052$\\
And~XI & $-6.3^{+0.6}_{-0.4}$ & $102^{+149}_{-1}$ & 0.85 & 51 & 72 & $0.415\pm0.069$\\
And~XII & $-7.0^{+0.7}_{-0.5}$ & $181^{+19}_{-87}$ & 0.43 & 9 & 19 & $0.321\pm0.123$\\
And~XIII & $-6.5^{+0.7}_{-0.5}$ & $115^{+207}_{-2}$ & 0.90 & 18 & 91 & $0.165\pm0.042$ \\
And~XIV & $-8.5^{+0.4}_{-0.3}$ & $161^{+81}_{-3}$ & 0.56 & 62 & 298 & $0.172\pm0.024$\\
And~XVII & $-7.8\pm0.3$ & $67^{+20}_{-24}$ & 0.82 & 136 & 180 & $0.430\pm0.044$ \\
And~XVIII & $-9.2^{+0.3}_{-0.4}$ & $457^{+39}_{-47}$ & 0.85 & 112 & 2346 & $0.046\pm0.004$ \\
And~XX & $-6.4^{+0.5}_{-0.4}$ & $128^{+28}_{-5}$ & 0.97 & 22 & 56 & $0.282\pm0.068$ \\
And~XXI & $-9.1\pm0.3$ & $135^{+8}_{-10}$ & 0.20 & 12 & 278 & $0.041\pm0.012$ \\
And~XXII & $-6.7^{+0.7}_{-0.5}$ & $275^{+8}_{-60}$ & 0.92 & 3 & 38 & $0.073\pm0.044$ \\
And~XXIII & $-9.8^{+0.2}_{-0.3}$ & $127^{+7}_{-4}$ & 0.19 & 8 & 382 & $0.021\pm0.007$ \\
And~XXIV & $-8.4\pm0.4$ & $169\pm29$ & 0.36 & 3 & 59 & $0.048\pm0.029$ \\
And~XXV & $-9.0\pm0.3$ & $90^{+57}_{-10}$ & 0.26 & 18 & 235 & $0.071\pm0.017$ \\
And~XXVI & $-5.8^{+0.9}_{-1.0}$ & $103^{+234}_{-3}$ & 0.87 & 19 & 71 & $0.211\pm0.053$ \\
And~XXIX & $-8.5\pm0.3$ & $198^{+18}_{-10}$ & 0.76 & 28 & 467 & $0.057\pm0.011$ \\
Cas~II & $-11.4\pm0.3$ & $145^{+95}_{-4}$ & 0.76 & 33 & 268 & $0.110\pm0.020$ \\
Cas~III & $-12.3\pm0.7$ & $144^{+6}_{-4}$ & 0.12 & 137 & 3147 & $0.042\pm0.004$ \\
Lac~I & $-11.4\pm0.3$ & $264\pm6$ & 0.28 & 110 & 3072 & $0.035\pm0.003$ \\
Per~I & $-10.3\pm0.7$ & $374^{+14}_{-10}$ & 0.71 & 10 & 806 & $0.012\pm0.004$ \\
\hline
And~XX field (WFC3) &  &  &  & 2 & 4 & $0.333\pm0.136$\\
And~XXI field (WFC3) &  &  &  & 1 & 5 & $0.167\pm0.068$\\ 
And~XXVI field (WFC3) &   &  &  & 0 & 5 & 0.0\\
\hline
\end{tabular}
\end{table*}

\acknowledgments
Based on observations made with the NASA/ESA Hubble Space Telescope, obtained [from the Data Archive] at the Space Telescope Science Institute, which is operated by the Association of Universities for Research in Astronomy, Inc., under NASA contract NAS 5-26555. These observations are associated with program HST-GO-11084, HST-SNAP-13442, and HST-GO-13699. Support for this work was provided by NASA through grant GO-13699 from the Space Telescope Science Institute, which is operated by AURA, Inc., under NASA contract NAS5-26555.



\end{document}